\documentclass[nofootinbib,aps,twocolumn,groupedaddress]{revtex4}  
\usepackage{graphicx} 
\usepackage{dcolumn} 
\usepackage{bm}        
\usepackage{amssymb}
\usepackage{physics}
\usepackage{ulem}

\def\beq{\begin{eqnarray}}
\def\eeq{\end{eqnarray}}

\def\({\left(}
\def\){\right)}

\def\mpl{M_{\rm pl}}

\newcommand{\be}{\begin{equation}}
\newcommand{\ee}{\end{equation}}
\newcommand{\la}{\langle}
\newcommand{\ra}{\rangle}
\def\ba{\begin{eqnarray}}
\def\ea{\end{eqnarray}}

\def\beq{\begin{eqnarray}}
\def\eeq{\end{eqnarray}}

\def\({\left(}
\def\){\right)}
\def\mn{_{\mu \nu}}

\def\p{\partial}
\def\la{\langle}
\def\ra{\rangle}

\def\lsim{\mathrel{\rlap{\lower3pt\hbox{\hskip0pt$\sim$}}
     \raise1pt\hbox{$<$}}}         
\def\gsim{\mathrel{\rlap{\lower4pt\hbox{\hskip1pt$\sim$}}
     \raise1pt\hbox{$>$}}}         

\def\lsim{\mathrel{\rlap{\lower3pt\hbox{\hskip0pt$\sim$}}
     \raise1pt\hbox{$<$}}}         
\def\gsim{\mathrel{\rlap{\lower4pt\hbox{\hskip1pt$\sim$}}
     \raise1pt\hbox{$>$}}}         

\usepackage{amsmath}
\usepackage{amsfonts}
\usepackage{verbatim}
\usepackage{graphicx}
\usepackage{subfigure}
\usepackage{amssymb}
\usepackage[T1]{fontenc}
\usepackage{slashed}
\usepackage{color}

\begin{document}

\title{De Sitter space as BRST invariant coherent state of gravitons}
\author{Lasha Berezhiani}
\author{Gia Dvali}
\author{Otari Sakhelashvili}
\address{ }
\address{Max-Planck-Institut f\"ur Physik, F\"ohringer Ring 6, 80805 M\"unchen, Germany}
\address{ }
\address{Arnold Sommerfeld Center, Ludwig-Maximilians-Universit\"at, Theresienstra{\ss}e 37, 80333 M\"unchen, Germany}

\begin{abstract}

The $S$-matrix formulation indicates that a consistent embedding of de Sitter state in quantum gravity is possible exclusively as an excited quantum state constructed on top of a valid $S$-matrix vacuum such as Minkowski. In the present paper we offer such a construction of de Sitter in the form of a coherent state of gravitons. Unlike previous realizations of this idea, we focus on BRST invariance as the guiding principle for physicality. In order to establish the universal rules of gauge consistency, we study the BRST-invariant construction of coherent states created by classical and quantum sources in QED and in linearized gravity. Introduction of $N$ copies of scalar matter coupled to gravity allows us to take a special double scaling limit, a so-called species limit, in which our construction of de Sitter becomes exact. In this limit, the irrelevant quantum gravitational effects vanish whereas the collective phenomena, such as Gibbons-Hawking radiation, are calculable.
\end{abstract}
\maketitle

\section{Introduction}

The classical General Relativity (GR) gives no a priory preference 
to any space-time metric. All solutions 
of GR are equally legitimate, provided they satisfy minimal 
consistency requirements, such as a valid causal structure. 
For example, there is no reason to give any advantage 
to the Minkowski space-time over a de Sitter one. 
 This prompts thinking that, upon quantization,  
 any valid classical background of GR can/must serve as a legitimate vacuum of the quantum theory.

  However, the $S$-matrix formulation of quantum gravity, which 
  in particular is organic to string theory,  gives a very different perspective \cite{Dvali:2020etd}. 
  The consistency and double-scaling argument show that having de Sitter 
  as a valid $S$-matrix vacuum, inevitably trivializes 
 the gravitational $S$ matrix. The reason is that the rigidity of geometry and the quantum coupling of gravitons (or closed strings) are controlled by one and the same parameter, the 
Planck mass, $\mpl$.  This simple but profound fact lies in very foundation of quantum gravity.   

  Thus, the $S$-martix tells us that the only possibility for embedding 
  de Sitter in quantum theory is by  treating it as an excited state 
  constructed on top of a valid $S$-matrix vacuum, such as 
  Minkowski \cite{Dvali:2020etd}.  
  
Originally, such a description has been applied to de Sitter and other cosmological 
space-times in \cite{Dvali:2011aa,Dvali:2012en, Dvali:2013eja,Dvali:2014gua,Berezhiani:2016grw,Dvali:2017eba}.
 In this approach, sometimes referred to as 
``corpuscular resolution'', the non-trivial space-time geometry is treated  as a state with high occupation number of gravitons, which is approximately classical.  The natural candidate for such a state is a coherent state.  

  Coherent states are thought to represent an adequate quantum description of classical systems. They were introduced in quantum field theory by Glauber \cite{Glauber:1963tx}. They also play important role in eliminate infrared divergencies by dressing asymptotic charged states with soft photons \cite{Chung:1965zza,Kibble:1968sfb,Kibble:1968oug,Kibble:1968npb,Kibble:1968lka,Kulish:1970ut}. Coherent states have been also used to formulate the quantum picture of solitons \cite{Dvali:2015jxa}.

   In previous formulations  \cite{Dvali:2013eja, Dvali:2014gua, Dvali:2017eba}  of de Sitter as a coherent state, the certain universal tendencies were observed.  
   The known semiclassical features of de Sitter are 
   recovered as a result of time-evolution of the coherent state. 
   For example, a famous Gibbons-Hawking radiation
   with temperature set by the Hubble $H$,  
   emerges as a result of an actual decay of the coherent state of 
   gravitons.  It therefore uncovers new features that 
   in ordinary treatment are not visible.  In particular, 
   as a result of backreaction, the de Sitter state gradually 
   looses coherence and evolves into a 
 self-entangled state.  This results in a full departure 
 from the classical description after the time-scale of half-decay. 
  In pure gravity, the corresponding time-scale, a so-called 
 quantum break-time, is $t_Q \sim \mpl^2/H^3$, and becomes 
 shorter in the presence of more degrees of freedom.
This phenomenon has important 
implications both fundamentally and observationally. 
For example, it limits the duration of any de Sitter  Hubble 
patch by $t_Q$, thereby eliminating the possibility of
an eternally inflating state. The significance of this statement for the efficient beginning of inflation was discussed in \cite{Berezhiani:2015ola}.
The $S$-matrix exclusion of 
de Sitter landscape also has important implications 
for the concept of ``naturalness'' in particle physics \cite{Dvali:2021kxt}. 
In this light, it is crucial to have a satisfactory 
understanding of quantum de Sitter state.

     In the present paper we shall investigate the legitimacy of de Sitter as a coherent state 
  from the point of view of  BRST symmetry  (see e.g. \cite{Weinberg:1996kr}).   
  Due to the fact that gravity is a gauge theory, it is imperative to have a rigorous understanding of the rules for building physical coherent states therein. In this work, we would like to present such a construction within the canonical framework of BRST quantization. This work will cover the analysis for the scalar electrodynamics and linearized gravity, uncovering peculiarities of the procedure and capitalising on physical implications.
  
  We shall give a BRST invariant formulation of de Sitter 
 coherent state in a quantum theory of massless spin-$2$, 
 with the (positive) cosmological constant source and 
 a coupling to a quantum scalar matter.
 We shall perform various consistency analysis of scaling 
 properties of de Sitter obtained in this way.    

 Next, we shall introduce the $N$ species of the scalar matter.  
 It has been shown \cite{Dvali:2021bsy} that this theory allows for a special double-scaling limit with $N \rightarrow \infty$, so-called ``species limit'', in which the 
 quantum gravitational processes simplify significantly. 
 All graviton non-linearities as well as their loop 
 contributions vanish as $1/N$.
 At the same time, the collective quantum gravitational effects, 
 such as Gibbons-Hawking radiation, are finite and explicitly calculable.  
 We observe, that  BRST invariant coherent state implemented in this setup correctly captures the quantum features of de Sitter such as the effects of Gibbons-Hawking particle creation. 
 
 Our results indicate that the program of treating de Sitter 
 as coherent state of gravitons built on Minkowski vacuum passes 
 the essential quantum consistency tests.

\section{Coherent States with Global Charge}

In the absence of gauge symmetry, the construction of coherent states is straightforward. They can be built around the vacuum of the theory by the action of the field displacement operator (see e.g.  \cite{Berezhiani:2020pbv,Berezhiani:2021gph}), written in terms of canonical degrees of freedom without invoking their asymptotic representation. In particular, the coherent state for a complex scalar field takes the following form
\beq
|C\ra=e^{-i\int d^3 x\left( \Phi_{\rm c}\hat{\Pi}-\Pi_{\rm c}\hat{\Phi}+h.c. \right)}|\Omega\ra\,,
\label{globalcoh}
\eeq
where $\Phi_{\rm c}(x)$ and $\Pi_{\rm c}(x)$ are the c-number functions of spatial coordinates, $\hat{\Phi}$ and $\hat{\Pi}$ stand for the field and canonical conjugate momentum operators respectively, while $|\Omega\ra$ denotes the vacuum (i.e. the lowest energy eigenstate of the Hamiltonian). This state is constructed in such a way that it satisfies the following initial conditions
\beq
&&\la C|\hat{\Phi}|C\ra (t=0)=\Phi_{\rm c}(x)\,,\\
&&\la C|\hat{\Pi}|C\ra (t=0)=\Pi_{\rm c}(x)\,,
\eeq
which follow solely from canonical commutation relations and the absence of the tadpoles in the vacuum. In the presence of the global $U(1)$ symmetry that shifts the phase of $\hat{\Phi}$, the expectation value of the corresponding charge in the state in question is conserved and given by
\beq
\la C| \hat{Q} |C \ra=i\int d^3x \left( \Pi_{ c}\Phi_{ c}-\Pi_{ c}^*\Phi_{ c}^*  \right)\equiv Q_{ c}\,.
\label{globalcharge}
\eeq
It must be pointed out that the coherent states are not the eigenstates of the global charge operator. 
 Rather, they are a superposition of states with different charges. In particular, using canonical commutation relations and taking into account that the vacuum is a zero-charge eigenstate, one is led to
\beq
&&\hat{Q}|C\ra=Q_{ c} |C\ra+e^{-i\int d^3 x\left( \Phi_{ c}\hat{\Pi}-\Pi_{ c}\hat{\Phi}+h.c. \right)}\times\nonumber\\
&&~~~~~~~~~~~~~~~~~\times  i\int d^3 x'\left( \Phi_{ c}\hat{\Pi}+\Pi_{ c}\hat{\Phi}-h.c. \right)|\Omega\ra\,.~~~
\label{notqeigen}
\eeq
The presence of the second term is the reason for the state not being an eigenstate of $Q$. This will have peculiar ramifications for gauge theories.

\section{Quantum Electrodynamics}

Upon gauging the $U(1)$ symmetry the construction of coherent states needs revision, on account of the consistency requirement that needs to be imposed on physical states. Our starting point is the BRST invariant formulation of scalar electrodynamics
\beq
\label{enmlag}
\mathcal{L}=-\frac14 \hat{F}_{\mu\nu}^2+|D_\mu\hat{\Phi}|^2-m^2|\hat{\Phi}|^2-\partial_\mu \hat{B}\hat{A}^\mu+\frac12\xi \hat{B}^2\nonumber\\
+\partial_\mu \hat{\bar c}\partial^\mu \hat{c}\,,\hskip 30pt
\eeq
with $\hat{F}\mn\equiv\partial_\mu \hat{A}_\nu-\partial_\nu \hat{A}_\mu$ and $D_\mu\hat{\Phi}\equiv\partial_\mu\hat{\Phi}-ig\hat{A}_\mu\hat{\Phi}$ as usual and $\hat{c},\hat{\bar{c}}$ being Fadeev-Popov ghosts that are anti-commuting Lorentz scalars.
 Throughout this work, the repeated covariant spacetime indices shall be contracted by the Minkowski metric $\eta\mn={\rm diag}(1,-1,-1,-1)$. The theory is invariant under celebrated BRST transformations	
\beq
&&\delta \hat{A}_\mu=\theta\partial_\mu \hat{c}\,, \quad \delta  \hat{\bar{c}}=\theta\hat{B}\,,\quad \delta \hat{c}=\delta\hat{B}=0\,, \nonumber\\
&&\delta \Phi=ig \theta\hat{c} \hat{\Phi}\,, \quad \delta\Phi^\dagger=-ig\theta\hat{c}\hat{\Phi}^\dagger\,,\label{brsttransform}
\eeq
where $\theta$ is a Grassmann number that serves as a parameter of transformation. The conserved charge associated with this symmetry, which is of utmost importance for the construction of the physical Hilbert space, takes the following form
\beq
\label{brstcharge}
\hat{Q}_B=\int d^3 x \left[\hat{c}\left( g\hat{\rho}-\partial_j \hat{E}_j\right)+\hat{B}\hat{\Pi}_{\bar{c}}+\partial_j\left( \hat{c}\hat{E_j} \right)\right]\,;
\eeq
with $\hat{\rho}\equiv i(\hat{\Phi}\hat{\Pi}-\hat{\Pi}^\dagger\hat{\Phi}^\dagger)$ representing the $U(1)$ charge density, $\hat{E}_j\equiv \hat{F}_{0j}$ being the electric field operator and $\hat{\Pi}_{\bar{c}}$ denoting the conjugate momentum of the corresponding ghost field.

As it is well known, there are consistency requirements on physical states in gauge theories. For example, Gauss' law informs us that charges must be dressed with a corresponding gauge field configuration \cite{Dirac:1955uv,Bagan:1999jf,Hirai:2019gio}, as there can exist no naked charges in nature. Within the adopted framework for quantization, the physical states carry vanishing BRST charge, together with vanishing ghost number. Although the photon states are effortless to construct, by the means of the field displacement operator that satisfies Gauss' law, the dressing fields are somewhat more peculiar to incorporate.  

The pure gauge field coherent state, devoid of ghosts, can be constructed as
\beq
|A\ra=e^{-i\hat{f}_A}|\Omega\ra\,,
\label{photoncoh}
\eeq
with
\beq
\hat{f}_A\equiv \int d^3 x\left( A_j^{ c}\hat{E}_j-E_j^{ c}\hat{A}_j+A_0^{ c}\hat{B}-B^{ c}\hat{A}_0 \right)\,,
\eeq
and the quantities carrying the label 'c' being the c-number functions that specify the initial field-configuration. In particular, initial conditions for one-point expectation values simply follow from canonical commutation relations and are given by
\beq
\la A | \hat{A}_\mu | A\ra=A_\mu^{ c}\,,  ~~\la A | \hat{E}_j | A\ra=E_j^{ c}\,,~~\la A | \hat{B} | A\ra=B^{ c}\,;~~~
\eeq
with other fields having vanishing initial expectation values. Taking into consideration that the proper vacuum must be annihilated by the BRST charge, the physicality condition $\hat{Q}_B|A\ra=0$ leads to the following constraint
\beq
\int d^3x \left(-\hat{c}\partial_j E^c_j+\hat{\Pi}_{\bar{c}} B^c+\partial_j\left( \hat{c}E^c_j\right) \right)|\Omega\ra=0\,.
\label{freephconst}
\eeq
This is an exact expression, for the derivation of which we have taken advantage of $e^{-i\hat{f}_A}$ being the field displacement operator at the initial time (which in turn follows from the Baker-Cambell-Hausdorff formula and equal-time commutation relations) and the conservation of $\hat{Q}_B$. The last (boundary) term of \eqref{freephconst} vanishes not only for localized electric field configurations, but for more general ones as well. We will see this by employing the momentum-space decomposition of the ghost operator $\hat{c}$, when relevant. The consistency condition \eqref{freephconst} is straightforwardly satisfied for
\beq
\partial_j E_j^{ c}=0\,, \qquad B^{ c}=0\,.
\eeq
In other words, the electric field must satisfy charge-free Gauss' law. The introduction of nontrivial $B^c$ seems possible, but leads to unnecessary complications and will not be considered in this work.

Moving on to the dressing field, a naive attempt for completing an electrically charged coherent state \eqref{globalcoh} into a physical one would consist of dressing it by an appropriate coherent gauge field. In other words, one could consider the charged coherent states to be of the form 
\beq
|A\ra\otimes |C\ra\,.
\eeq

The inadequacy of this proposal can be straightforwardly demonstrated by showing that
\beq
\label{brstcharge}
\hat{Q}_B\{|A\ra\otimes |C\ra\}=\hat{Q}_Be^{-i\int d^3 x\left( A_j^{ c}\hat{E}_j-E_j^{ c}\hat{A}_j+A_0^{ c}\hat{B}-B^{ c}\hat{A}_0 \right)}\nonumber\\\times e^{-i\int d^3 x\left( \Phi_{ c}\hat{\Pi}-\Pi_{ c}\hat{\Phi}+h.c. \right)}|\Omega\ra\neq 0\,,~~
\eeq
as long as $\Phi_{ c}$ and $\Pi_{ c}$ are non-vanishing. The inability to dress the scalar coherent state with a coherent electromagnetic configuration is connected to the fact that the former is not an eigenstate of $U(1)$ charge; see e.g. \eqref{notqeigen}. Therefore, the matter states cannot be made BRST invariant by dressing them with coherent gauge field configuration. Instead, we begin the construction from the invariant operators. 

The idea of defining gauge invariant matter degrees of freedom goes back to Dirac \cite{Dirac:1955uv} (see also \cite{Bagan:1999jf}), according to which the operator undergoing merely a phase rotation under gauge transformations can be combined with the gauge field in the following invariant fashion
\beq
\label{u1invphi}
&&\hat{\Phi}_{ g}=\hat{\Phi}\cdot{\rm exp}\left(-ig\frac{1}{\laplacian}\partial_j \hat{A}_j\right)\,,\\
&&\hat{\Pi}_{ g}=\hat{\Pi}\cdot{\rm exp}\left(+ig\frac{1}{\laplacian}\partial_j \hat{A}_j\right)\,;
\label{u1invpi}
\eeq
with the subscript 'g' indicating the gauge invariance. 
Similar nonlocal operators ${\rm exp}(-ig\frac{1}{\Box}\partial_{\mu}A^{\mu})$ were used for maintaining gauge invariance 
in case of anomalous symmetries \cite{Dvali:2009ne}.

It is important to notice that these operators satisfy canonical commutation relations
\beq
[\hat{\Phi}_{ g}(t,x),\hat{\Pi}_{ g}(t,y)]=i\delta^{(3)}(x-y)\,.
\eeq
Moreover, it is straightforward to show explicitly that due to the outlined gauge invariant construction
\beq
\label{qcom}
[\hat{Q}_B,\hat{\Phi}_g(x)]=[\hat{Q}_B,\hat{\Pi}_g(x)]=0\,.
\eeq
Based on this observation, we can construct dressed coherent states for matter fields in analogy with \eqref{globalcoh}, albeit with $\hat{\Phi}_g$ and $\hat{\Pi}_g$. In other words, the coherent state
\beq
\label{qchargecoh}
|C_g\ra=e^{-i\int d^3 x\left( \Phi_{ c}\hat{\Pi}_g-\Pi_{ c}\hat{\Phi}_g+h.c. \right)}|\Omega\ra\,,
\eeq
satisfies the physicality conditions for arbitrary $ \Phi_{ c}(x)$ and $\Pi_{ c}(x)$. As for connection with the classical field configurations, it is straightforward to show that at the initial moment
\beq
\label{scalarinit}
\la C_g| \hat{\Phi}_g |C_g \ra (t=0)=\Phi_c\,, \quad \la C_g| \hat{\Pi}_g |C_g \ra (t=0)=\Pi_c\,,~~\\
\label{gaussinit}
\partial_j\la C_g| \hat{E}_j |C_g \ra (t=0)=ig\left(\Phi_c\Pi_c-\Pi^*_c\Phi^*_c\right)\,,\hskip 48pt\\
\label{Ainit}
\la C_g| \hat{A}_\mu |C_g \ra (t=0)=0\,.\hskip 133pt
\eeq

The state can be further supplemented with the coherent photon configuration using the gauge field displacement operator discussed above. Putting all the ingredients together, a physical coherent state that corresponds to a certain classical charge distribution and accounts for the dressing field as well as for the additional electromagnetic field is given by
\beq
\label{genst}
&&|C_g,A\ra=e^{-i\int d^3 x\left( \Phi_{ c}\hat{\Pi}_g-\Pi_{ c}\hat{\Phi}_g+h.c. \right)}\nonumber\\
&&\hskip 45pt\times e^{-i\int d^3 x\left( A_j^{ c}\hat{E}_j-E_j^{ c}\hat{A}_j+A_0^{ c}\hat{B} \right)}|\Omega\ra\,.
\eeq
It must be stressed that the order in which the exponentials appear in this definition of the coherent state is necessary for reproducing \eqref{scalarinit}. The consistency of this state requires the transversality of $E_j^c$, thus \eqref{gaussinit} will be maintained; the only modification being the generation of the nonvanishing expectation value for $\hat{A}_\mu$.

The dynamics can be readily obtained from Heisenberg's equations that follow from the Hamiltonian of the system
\beq
\label{hamilton}
\hat{H}=\int d^3 x\left[ \frac{1}{2}\hat{E}_j^2+\frac{1}{4}\hat{F}_{ij}^2+|\hat{\Pi}|^2+|D_j\hat{\Phi}|^2+m^2|\hat{\Phi}|^2\right. \nonumber\\
+\hat{B}\p_j\hat{A}_j-\frac{\xi}{2}\hat{B}^2 +\p_j \left(\hat{A}_0\hat{E}_j-\hat{B}\hat{A}_j\right)\nonumber\\
+\hat{A}_0 \left( -\p_j \hat{E}_j+ig\left( \hat{\Phi}\hat{\Pi}-\hat{\Phi}^\dagger\hat{\Pi}^\dagger \right) \right)\nonumber\\
+\hat{\Pi}_c\hat{\Pi}_{\bar c}+\p_j \hat{\bar{c}}~\p_j\hat{c}\Big]\,.~
\eeq

Heisenberg's operator equations for the gauge sector lead to the following set of equations for 1-point functions
\begin{flalign}
\label{A0eq}
&\p_0 \la\hat{A}_0\ra=\p_j\la \hat{A}_j \ra-\xi \la\hat{B}\ra\,,\\
\label{Ajeq}
&\p_0 \la \hat{A}_j\ra=\la\hat{E}_j\ra+\p_j\la\hat{A}_0\ra\,,\\
\label{Eeq}
&\p_0 \la\hat{E}_{j}\ra-\p_i \la\hat{F}_{i j}\ra-\p_j \la\hat{B}\ra=ig\la \hat{\Phi}^\dagger\hat{D}_j \hat{\Phi}-h.c. \ra\,,~~\\
\label{Beq}
&\p_0 \la\hat{B}\ra=-\p_j \la\hat{E}_{j}\ra+ig\la \hat{\Phi}\hat{\Pi}-\hat{\Pi}^\dagger\hat{\Phi}^\dagger \ra\,,\end{flalign}
with $\la\ldots \ra$ denoting the expectation value of the enclosed Heisenberg picture operator in the coherent state \eqref{genst}. These equations represent quantum extensions of classical equations of motion that follow from \eqref{enmlag}. The nonlinear terms of \eqref{Eeq} and \eqref{Beq}, once evaluated over a coherent state, will contain both classical and quantum contributions. The latter can be quantified by
\beq
\label{sj}
&&S_j\equiv i\la \hat{\Phi}^\dagger\hat{D}_j \hat{\Phi}-(\hat{D}_j \hat{\Phi})^\dagger\hat{\Phi}\ra-\bar{J}_j \,,\\
&&S_0\equiv i\la \hat{\Phi}\hat{\Pi}-\hat{\Pi}^\dagger\hat{\Phi}^\dagger \ra-\bar{\rho}\,,
\label{s0}
\eeq
where $\bar{J}_\mu$ has been defined as the current constructed merely out of 1-point functions, i.e. 
\beq
&&(-i)\bar{J}_j\equiv \la \hat{\Phi}^\dagger\ra \left(\p_j \la\hat{\Phi}\ra-ig\la A_j\ra\la\hat{\Phi}\ra\right)-h.c.\,, \\
&& (-i)\bar{\rho}\equiv \la\hat{\Phi}\ra\la\hat{\Pi}\ra-\la\hat{\Pi}^\dagger\ra\la\hat{\Phi}^\dagger \ra\,.
\eeq

As a result, \eqref{Eeq} and \eqref{Beq} become
\beq
\label{1pointE}
&&\p_0 \la\hat{E}_{j}\ra-\p_i \la\hat{F}_{i j}\ra-\p_j \la\hat{B}\ra=g\bar{J}_j+gS_j\,,\\
\label{1pointB}
&&\p_0 \la\hat{B}\ra=-\p_j \la\hat{E}_{j}\ra+g\bar{\rho}+gS_0\,,
\eeq
In the absence of $S_0$ and $S_j$ these are classical equations of motions for the theory at hand. The computation of these quantum terms requires the knowledge of 2-point and 3-point functions, however we could instead  follow \cite{Berezhiani:2021gph} evaluating them explicitly in the coherent state up to a desirable order in $\hbar$ and $g$. 
The explicit form of the equation for the scalar field will be useful when discussing the particle production and will be given when relevant.
For now, let us keep in mind that the coherent state sets the stage by providing initial conditions for \eqref{A0eq},  \eqref{Ajeq}, \eqref{1pointE}, \eqref{1pointB} and for the scalar field.

\section{Classical Charges}

In this section we would like to discuss the possibility of introducing fundamentally classical sources as electromagnetic analog of the cosmological constant (we borrow this analogy from 
 \cite{Dvali:2007kt}). We proceed by adding the following term to the Lagrangian
\beq
\Delta \mathcal{L}=-\hat{A}_\mu J^\mu_{\rm cl}\,,\quad \text{with}\quad \p_\mu J^\mu_{\rm cl}=0\,;
\eeq
where $J^\mu_{\rm cl}$ is a four-vector of predetermined c-number functions. Its presence does not alter the BRST transformation properties \eqref{brsttransform}, under which the Lagrangian density is invariant up to a total derivative 
\beq
\delta \left(\mathcal{L}+\Delta\mathcal{L}\right)=-\p_\mu\left( \theta\hat{c} J^\mu_{\rm cl} \right)\,.
\eeq
As a result the Noether charge needs to be amended correspondingly, resulting in
\beq
\label{brstj}
\hat{Q}^J_B=\hat{Q}_B+\int d^3 x~ \hat{c}J^0_{\rm cl}\,,
\eeq
with $\hat{Q}_B$ denoting the BRST charge in the absence of the classical source and is given by \eqref{brstcharge}.

In the classical limit, the presence of $J_{\rm cl}$ would source classical electromagnetic field configuration. One could be tempted to associate this state to a vacuum of the theory.
However, due to the fact that such a state is expected to precipitate the particle production, the vacuum treatment 
is legitimate only in the limit of zero backreaction \cite{Dvali:2020etd}.  
In gravity,  where the 
existence of valid $S$-matrix vacuum is vital, the analogous issue has profound consequences 
for positive cosmological constant. Namely, it demands 
that de Sitter space be treated as coherent state built around the Minkowski vacuum.
  For the theory at hand, this corresponds to constructing the coherent state around the vacuum of the Hamiltonian \eqref{hamilton}.\footnote{In different contexts, the coherent state of electromagnetic field of classical charges  has been discussed previously (see, e.g., \cite{Muck:2013orm,Muck:2015dea}).}

Due to the fact that $\hat{Q}^J_B$ commutes with the total Hamiltonian
\beq
\hat{H}_J=\hat{H}+\int d^3 x \hat{A}_\mu J_{\rm cl}^\mu\,
\eeq
we would like to construct a coherent state of electromagnetic field which satisfies
\beq
\hat{Q}^J_B|J\ra=0\,.
\label{constrj}
\eeq
As per the arguments given above, we would like to use the vacuum $|\Omega\ra$ of $\hat{H}$ as the basis. It is straightforward to show that the state
\beq
\label{jcoh}
|J\ra=e^{-i\int d^3 x\left( A_j^{ c}\hat{E}_j-E_j^{ c}\hat{A}_j+A_0^{ c}\hat{B} \right)}|\Omega\ra
\eeq
constructed in analogy with the pure photon state, satisfies the required constraint \eqref{constrj} if
\beq
\int d^3x \left[\hat{c}\left(-\partial_j E^c_j+J^0_{\rm cl}\right)+\partial_j\left( \hat{c}E^c_j\right) \right]|\Omega\ra=0\,.
\label{jclconst}
\eeq
This straightforwardly entails Gauss' law
\beq
\p_j E_j^c=J^0_{\rm cl}\,,
\label{GLcl}
\eeq
as long as the boundary term of \eqref{jclconst} vanishes. The latter is trivially satisfied by localized classical sources, while its applicability to more general cases will be considered in the next section.

It must be stressed that $|\Omega\ra$, being the vacuum of $\hat{H}$, is annihilated by $\hat{Q}_B$ and not by $\hat{Q}^J_B$. Consequently, the physicality condition satisfied by it is not preserved by the Hamiltonian flow generated by $\hat{H}_J$.

Next, we would like to ask if the coherent state $|J\ra$ has a consistent dynamics. The evolution of 1-point expectation values are governed by the equations similar to the ones derived in the previous section, albeit with additional classical source on the right hand side of \eqref{Eeq} and \eqref{Beq}. The process of interest is the Schwinger pair production of $\Phi\Phi^\dagger$ and the subsequent backreaction on the 1-point function of the electric field. It is straightforward to see that to the leading order the evolution of the quantum terms is determined by the time-dependence of the tree-level mode functions for the scalar field, which in turn is governed by the electromagnetic background field. Moreover, the origin of the latter is immaterial for the former, since the tree-level scalar mode functions are only sensitive to the classical background of the vector field. Therefore, as long as charges are lighter than the pair-creation threshold value, the Schwinger pair production will begin and the backreaction on the electromagnetic field-background will be governed by \eqref{A0eq}-\eqref{Beq} supplemented with the external source. The process will continue until the produced particles
decrease the electric field below the Schwinger threshold. 

\section{Infinitely Homogeneous Classical Source}

Let us finish the discussion of quantum electrodynamics (QED) by focusing on a homogeneous constant source filling the entire space.  This represents the closest electromagnetic analog of gravity with cosmological constant  \cite{Dvali:2007kt}.  

The toy model consists of 
the scalar electrodynamics supplemented with homogeneous external current $J_{\rm cl}^\mu=\delta^\mu_0 \rho=const$. The coherent electromagnetic state \eqref{jcoh} sourced by this current must satisfy the physicality condition \eqref{jclconst}. As we saw, this condition straightforwardly leads to Gauss' law for the electric field configuration $E_j^c$ parameterizing the coherent state produced by localized classical sources, for which the boundary term of \eqref{jclconst} vanishes trivially. 
Let us now assess what happens to it for the homogeneous charge distribution. 

The electric field produced by a constant charge is a 
linear function of coordinates.  Such a field grows unbounded 
towards infinity. In the presence of any sort of dynamical charges 
in the theory, the field will be subjected to a discharge.
It is therefore clear that a classical solution of a linearly growing 
field  cannot be sustained in quantum theory. 
 Nevertheless, it is useful to consider a regularized version 
 of the story with imposed spherical boundary which is 
 gradually  taken to infinity.

According to Gauss' law \eqref{GLcl}, $E_j^c$ does no longer vanish on the boundary as it is given by
\beq
E_j^c=\frac{\rho}{3} x_j\,;
\label{constrhoE}
\eeq
in fact, this quantity diverges at large distances. We can nevertheless demonstrate that the above mentioned boundary term vanishes. For this, it suffices to notice that 
without loss of generality
at any given moment of time the ghost field can be expanded  in creation-annihilation operators
\beq
\hat{c}(x,t_0)=\int \frac{d^3 k}{(2\pi)^3}\frac{1}{\sqrt{2k}}\left( \hat{a}_k e^{i \vec{k}\cdot\vec{x}}+\hat{b}^\dagger_k e^{-i \vec{k}\cdot\vec{x}} \right)\,.
\eeq
If we further take into account that the vacuum is annihilated by $\hat{a}_k$, we are led to the following equivalence
\beq
\int d^3x~ \partial_j( \hat{c}E^c_j)|\Omega\ra=0\, \Leftrightarrow \int d^3x~ \partial_j\left( e^{-i \vec{k}\cdot\vec{x}}x_j\right)=0\,.~
\eeq
It is straightforward to see that the latter simplifies to
\beq
\frac{\partial}{\partial k_j}\left( k_j\delta^{(3)}(k) \right)=0\,,
\eeq
which holds as one of the properties of Dirac's delta-function. Therefore, the coherent state of the electromagnetic field, produced  by the external source in question,  is consistent with the physicality conditions of BRST quantization. It must be stressed that this does not fully prove 
the legitimacy of the construction, since the exact classicality of the source can lead to other inconsistencies such as unboundedness of the Hamiltonian from below. We shall set aside such issues and instead discuss the dynamical aspects of the constructed BRST-invariant coherent state.

Simplifying the construction, we take $A_\mu^c=0$ in \eqref{jcoh}, which corresponds to $\la \hat{A}_\mu\ra=0$ at the initial time, leaving us with the following coherent state
\beq
\label{rhocoh}
|\rho\ra=e^{i\int d^3 x\left( E_j^{ c}\hat{A}_j \right)}|\Omega\ra\,.
\eeq
Next we are interested in computing the leading quantum corrections to the dynamics of 1-point expectation values for the gauge sector, resulting from the quantum pair creation of scalar particles. As it was demonstrated explicitly in \cite{Berezhiani:2021gph}, the relevant dynamics follows from \eqref{A0eq}-\eqref{Beq} (albeit with additional classical source) by retaining only the expectation values of bilinear operators in $S$. For the setup in question, the equations take the following form
\beq
\label{A0cl1}
\p_0 \la\hat{A}_0\ra=\p_j\la \hat{A}_j \ra-\xi \la\hat{B}\ra\,,\hskip 110pt\\
\label{Ecl1}
\p_0 \la\hat{E}_{j}\ra-\p_i \la\hat{F}_{i j}\ra-\p_j \la\hat{B}\ra=ig\la \rho | \hat{\Phi}^\dagger D^{\rm cl}_j\hat{\Phi}-h.c. |\rho\ra\,,\\
\label{Bcl1}
\p_0 \la\hat{B}\ra+\p_j \la\hat{E}_{j}\ra-\rho=ig\la \rho|\hat{\Phi}\hat{\Pi}-h.c. |\rho\ra\,,\hskip 45pt
\eeq
where $D^{\rm cl}_\mu\equiv \p_\mu-igA_\mu^{\rm cl}$.
The terms on the right hand side of \eqref{Ecl1} and \eqref{Bcl1} are already quantum, since 2-point functions are of order $\hbar$ in the leading order. Therefore, we need to find the relevant correlators at the tree level. The required dynamics of $\hat{\Phi}$ follows from solving the following semiclassical equation
\beq
\left(D^{\rm cl}_\mu D_{\rm cl}^\mu +m^2\right)\hat{\Phi}(x,t)=0,
\label{phiprod}
\eeq
with the classical electromagnetic configuration for our setup given by $A_j^{\rm cl}=\frac{1}{3}\rho tx_j$, $A_0^{\rm cl}=\frac{1}{2}\rho t^2$ and $B^{\rm cl}=0$.
This equation needs to be solved with appropriate initial conditions, which for our coherent state corresponds to
\beq
\la \rho |\hat{\Phi}(x,0)\hat{\Phi}(y,0) |\rho \ra=\la \Omega |\hat{\Phi}(x,0)\hat{\Phi}(y,0) |\Omega \ra\,.
\eeq
In other words, even though the dynamics is given by the background dependent equation of motion, the initial conditions for the mode functions are the ones in Minkowski vacuum. See \cite{Berezhiani:2020pbv,Berezhiani:2021gph} for the discussion of this point and implications for perturbative dynamics.

The time dependence of the background field appearing in \eqref{phiprod} will facilitate the particle production as expected.

 In the process, the 1-point function of the scalar field will remain zero, with nontrivial dynamics showing up in the 2-point correlation function. The latter will backreact on dynamics of 1-point functions of the gauge sector as per \eqref{Ecl1} and \eqref{Bcl1}. Obviously, there will be divergences among the quantum terms, an obvious one emerging due to the appearance of the equal-point 2-point function, which can be renomalized using standard prescriptions. The leftover finite quantum correction will amount to the physical backreaction.

There is a finite radius at which the field strength will cross the Schwinger pair creation threshold.  Beyond it the efficient production of charges will take place until the background electric field is sufficiently reduced. This crossover radius can be estimated as
\beq
R\sim \frac{m^2}{g\rho}\,.
\eeq
This radius marks the boundary of validity 
of the semiclassical approximation. 
Beyond it a uniform charge distribution 
undergoes a rapid quantum breaking.

\section{Linear Gravity}

Next we would like to understand the ramifications of the BRST constraint for gravitational systems and specifically for the de Sitter space. We shall primarily work within linearized gravity, followed by the introduction of interactions with the matter perturbatively. 

The original construction of de Sitter coherent state \cite{Dvali:2013eja, Dvali:2017eba} was based on a  correspondence between the 
classical de Sitter metric of linear Einstein theory and the 
source-free solution of massive Fierz-Pauli theory.  This
classical correspondence has been established earlier
in \cite{Dvali:2007kt}. The 
idea \cite{Dvali:2013eja, Dvali:2017eba} was to map the de Sitter space sourced by the cosmological constant on the solution of 
Fierz-Pauli theory, with subsequent quantum resolution   
of the latter. This allowed to construct de Sitter 
as a quantum state of high occupation number of 
nearly on-shell Fierz-Pauli gravitons on Minkowski. 
This accomplishes a first necessary step towards realization of   
 de Sitter state in $S$-matrix formulation   
of quantum gravity. In this framework,  
the processes such as Gibbons-Hawking particle creation are represented with $S$-matrix process of decay of the graviton coherent state. At finite $\mpl$ this leads to backreactions, 
such as loss of coherence. 
In this approach, the consistency of Hilbert space is guaranteed
by gauge invariance of Fierz-Pauli gravitons.
This invariance is due to 
additional polarizations of massive graviton
(as compared to massless case) which 
play the role of St\"uckelberg fields that maintain the gauge 
invariance of the graviton state. 
Such invariance, of course, is not experienced by
 the states of massless Einstein gravitons.  
 Therefore, a construction of consistent de Sitter coherent state 
 directly in linear Einstein requires additional measures. 
 Such measures will be developed in the present 
 work in the form of  BRST invariance of the state.

Our starting point is the BRST invariant formulation of the theory of  a free spin-2 field in Minkowski space,
\beq
\mathcal{L}=\frac12 (\partial_\alpha \hat h_{\mu\nu})^2-\frac12(\partial_\alpha \hat h)^2+\partial_\alpha \hat h \partial_\mu \hat h^{\mu\alpha}
-\partial_\mu \hat h^{\mu\alpha}\partial_\nu \hat h^{\nu}_\alpha\nonumber \\
-\partial_\mu \hat B_\nu\left( \hat h^{\mu\nu}-\frac{1}{2}\eta^{\mu\nu} \hat h\right)+\frac12\xi \hat B_\mu^2+\partial_\alpha\hat{\bar{C}}_\mu\partial^\alpha \hat C^\mu\,,~~
\label{freeh}
\eeq
which is the linearized version of the framework developed in \cite{Kugo:1978rj}. It follows from Einstein's theory of gravity in the limit of infinite Planck's mass. Here $\hat h_{\mu\nu}$, $\hat C_\mu$ and $\hat B_\mu$ are the massless spin-$2$ (graviton) field, Fadeev-Popov ghost vectors and an auxiliary vector respectively.  $\xi$ is the gauge fixing parameter. The theory at hand is invariant under the BRST transformation
\begin{eqnarray}
\label{hbrst}
&&\delta \hat h_{\mu\nu}=\theta\left( \partial_\mu \hat C_\nu+\partial_\nu \hat C_\mu\right)\,,\\
&&\delta \hat{\bar{C}}_\mu=\theta \hat B_\mu\,,
\label{Cbrst}
\end{eqnarray}
where $\theta$ is a Grassmann variable, with the rest of the fields transforming trivially.

In analogy with the previous section, the construction of states is performed within the canonical Hamiltonian framework. Following the ADM formalism, we supplement the Lagrangian \eqref{freeh} with  boundary terms appropriate for removing the time-derivatives of $\hat{h}_{00}$ and $\hat{h}_{0j}$. As a result, the conjugate momentum of $\hat{h}_{ij}$ reduces to
\beq
\hat{\pi}_{ij}=\p_0 h_{ij}-\delta_{ij}\p_0h_{kk}+2 \delta_{ij}\p_k h_{k0}-\p_i h_{j0}-\p_j h_{i0}\,.~~
\eeq
Its BRST transformation property follows from \eqref{hbrst} and is given by
\beq
\label{pibrst}
\delta \hat{\pi}_{ij}=2\theta\left( \laplacian \delta_{ij}-\p_i\p_j \right)\hat{C}_0\,.
\eeq

Following the footsteps of our QED consideration, we construct the physical states in a BRST invariant fashion. A  pure-graviton coherent state, free of ghosts, can be built as
\beq
|h\ra=e^{-i\int d^3 x\left( h_{ij}^{ c}\hat{\pi}_{ij}-\pi_{ij}^{ c}\hat{h}_{ij}+B^{ c}_\mu\hat{\Pi}^{\mu}-\Pi_c^{\mu}\hat{B}_\mu \right)}|\Omega\ra\,,
\label{gravitoncoh}
\\
\Pi^\mu\equiv -h^{0\mu}+\frac{1}{2} \eta^{0\mu}h\,,\nonumber
\eeq
with c-number functions setting the initial expectation values of corresponding operators and $|\Omega\ra$ denoting the Minkowski vacuum. In analogy with QED, 
from the beginning,  we shall take $B^{ c}_\mu=0$, as its presence would introduce an unnecessary complication in BRST condition which is easily satisfied otherwise.
 Imposing BRST invariance we are immediately led to
\beq
\hat Q_B  \ket{h}= e^{-i\int d^3 x\left( h_{ij}^{ c}\hat{\pi}_{ij}-\pi_{ij}^{ c}\hat{h}_{ij}-\Pi^{\mu}_{ c}\hat{B}_\mu\right)} \hskip 60pt \nonumber\\ 
\times\int d^3 x\left(-2h_{ij}^c(\laplacian\delta_{ij}-\partial_i\partial_j)\hat C_0+2\pi_{ij}^c\partial_i\hat C_j\right)|\Omega\ra\,.~
\eeq
The consistency of the state requires the above expression to vanish. Upon integrating the second line by parts we are led to the relations reminiscent of the classical constraints of linear gravity 
\beq
\label{classconst}
\left(\laplacian\delta_{ij}-\partial_i\partial_j\right)h_{ij}^c=0\,,\quad \text{and} \quad \partial_i\pi_{ij}^c=0\,,
\eeq
that need to be satisfied by the physical configuration along with the following boundary conditions
\begin{eqnarray}
\label{boundary1}
&&	\int d^3x ~\partial_i \left(\pi^c_{ij}\hat C_j\right)|\Omega\ra=0\,,\\
\label{boundary2}
&&	\int d^3x~\partial_i \left(\left[\partial_j \hat C_0-\hat C_0 \partial_j\right] (\delta_{ij}h_{kk}^c-h_{ij}^c)\right)|\Omega\ra=0\,.~~~~~~	\label{boundary2}
\end{eqnarray}
Notice that $h_{0\mu}^c$ is unrestricted, due to the fact that the corresponding operators represent Lagrange multipliers, similar to $A_0$ in electrodynamics. 
 A further parallel can be drawn with QED, by pointing out that \eqref{classconst} represents the gravitational counterpart to the charge-free Gauss' law, while \eqref{boundary1} and \eqref{boundary2} are equivalent to the boundary term of \eqref{freephconst} and are automatically satisfied for configurations that vanish on the boundary.

Having demonstrated how to construct physical coherent states of gravitons over the Minkowski vacuum in a source-free theory at hand \eqref{freeh}, we would like to begin introducing sources. As it has been already mentioned, the case of particular interest is the cosmological constant. The latter is a classical source of gravity that is incorporated in a linear theory by adding the following term to the Lagrangian
\begin{equation}
	\Delta\mathcal{L}=-\lambda\hat h\,.
	\label{lincc}
\end{equation}

Obviously, this addition does not alter the expressions for the canonical momenta. Nevertheless, since it is invariant under \eqref{hbrst} only up to a total derivative, it gives the additional contribution to the Noether charge
\beq
\label{fullbrst}
\hat{Q}_B^\lambda=\hat{Q}_B+\int d^3 x 2\lambda \hat C_0\,.
\eeq
This is the charge with respect to which we must define the physical Hilbert space, since it commutes with the Hamiltonian of the system in the presence of the cosmological constant. That is,  $[\hat{H},\hat{Q}_B^\lambda]=0$ while $[\hat{H},\hat{Q}_B]\neq 0$. Here $\hat{Q}_B$ is the BRST charge in the absence of the cosmological constant, in complete analogy with our consideration of classical charges in QED. Notice that we have not provided the explicit expression for $\hat{Q}_B$ because of lack of necessity.

A classical solution of linearized equation with cosmological constant source is given by  \cite{Dvali:2007kt}
\beq
h_{ij}=-\frac{\lambda}{6}\left( t^2\delta_{ij}+x_i x_j \right)\,,\quad h_{00}=h_{0j}=0\,,
\label{hcl}
\eeq
which, for $\lambda >0$, represents the short-scale approximation of the de Sitter spacetime in closed slicing. In fact, switching to the dimensionless metric $g\mn=\eta\mn+h\mn/\mpl$, it is straightforward to see that \eqref{hcl} follows from the latter in $\mpl\rightarrow \infty$ limit while $\lambda$ is held fixed.  So does \eqref{freeh} supplemented with \eqref{lincc} from the fully nonlinear Einstein-Hilbert action with the cosmological constant. It must be noted that in this limit the curvature of the spacetime vanishes,
\beq \label{Hubble} 
H^2\simeq \frac{\lambda}{\mpl}\rightarrow 0\,.
\eeq
This is equivalent to taking the Hubble radius to infinity. Thus, \eqref{hcl} is a good approximation of the de Sitter spacetime only at deeply sub-horizon scales.

As already commented, in the original proposal \cite{Dvali:2013eja, Dvali:2017eba}, the quantum resolution 
of the metric (\ref{hcl}) as of multi-graviton coherent state on 
Minkowski was 
achieved via representing the Einstein graviton as 
a component of Fierz-Pauli field. The gauge invariance 
of the latter served as assurance for the consistency of such a state. Here we take a different path.

The question we ask is whether the de Sitter background 
(\ref{hcl}) can be represented as a BRST-invariant coherent state of gravitons built over the Minkowski vacuum. Let us begin by noticing that, due to the presence of the cosmological constant, the Minkowski vacuum $|\Omega\ra$ is no longer annihilated by the full BRST charge given by  \eqref{fullbrst}, but only by $\hat{Q}_B$ which no longer commutes with the Hamiltonian. The similar observation was made for quantum electrodynamics in the presence of classical charges. There it was nevertheless possible to construct a BRST-invariant coherent state over the vacuum of the theory free of external classical sources. To demonstrate the same for the linear gravity, we move forward by constructing the coherent state corresponding to \eqref{hcl} at $t=0$ as
\beq \label{CohH}
|h\ra=e^{-i\int d^3 x~\left(h_{ij}^{ c}\hat{\pi}_{ij}+\frac{1}{2}h_{kk}^c \hat{B}_0\right)}|\Omega\ra\,,\\
 \text{with}\quad h_{ij}^{ c}=-\frac{\lambda}{6}x_i x_j\,.
\eeq

It is straightforward to see that in the presence of the cosmological constant the first equation of \eqref{classconst} is modified into
\beq
\left(\laplacian\delta_{ij}-\partial_i\partial_j\right)h_{ij}^c-\lambda=0\,.
\label{lambdaconst}
\eeq
This is the direct consequence of the above-mentioned point about $|\Omega\ra$ not being annihilated by $\hat{Q}^\lambda_B$. The constraint \eqref{lambdaconst} is readily satisfied by \eqref{hcl}. The boundary conditions \eqref{boundary1} and \eqref{boundary2} persist without modification, the first of which is trivially satisfied for the configuration at hand. The demonstration for the second one, on the other hand, requires expansion of $\hat{C}_0(x,t)$ in ladder operators, as it was done for electrodynamics in the previous section. It is nevertheless straightforward to show that it holds.

\section{Scalar matter coupled to gravity}
\label{scalarsection}

  As the next step, we shall introduce coupling of graviton 
 to a scalar field.  
Truncating the theory of a scalar field coupled to Einstein gravity at the first nontrivial order in $\mpl^{-1}$ expansion, we arrive at 
\begin{equation}
\label{pertL}
	\mathcal{L}=\mathcal{L}_\phi+\mathcal{L}_h-\frac{1}{2\mpl}\hat{h}_{\mu\nu}\hat{T}^{\mu\nu}+\mathcal{O}(\mpl^{-2})\,,
\end{equation}
where  $\mathcal{L}_\phi$ is a Lagrangian of the scalar field in the absence of gravity, $\mathcal{L}_h$ is given by \eqref{freeh} describing the free propagation of the graviton (albeit with additional nonlinear gauge fixing and ghost parts \cite{Kugo:1978rj}) and $\hat{T}_{\mu\nu}$ stands for the energy-momentum tensor of the first two terms of \eqref{pertL}. The graviton contribution to $T\mn$ warrants the modification of graviton's BRST transformation property \eqref{hbrst} to include the ``non-Abelian'' correction. However, the latter will not contribute to the effects of order $\mpl^{-1}$ that we are after, as it will become clear shortly. The transformation property of the scalar field is given by
\begin{equation}
\label{dphigrav}
	\delta \hat\phi=\frac{\theta}{\mpl}\hat C^\mu\partial_\mu\hat\phi\,,
\end{equation}
which takes the following form when rewritten in terms of canonical variables 
\beq
\delta\hat \phi=\frac{\theta}{\mpl}\left(\hat\Pi_\phi \hat C_0-\partial_j\hat\phi \hat C_j\right)+\mathcal{O}(\mpl^{-2})\,
\eeq
(Assuming canonical kinetic term for the scalar. The generalization is straightforward and will not be pursued here).

In analogy with electrodynamics, the BRST invariant dressing of the scalar coherent state amounts to replacing the scalar field operators by their invariant counterparts. For example, what serves for the invariant version of $\hat{\phi}(x^\mu)$ is simply $\hat{\phi}(x^\mu+\epsilon^\mu)$, with $\epsilon$ being a function of the graviton field that transforms as
\beq
\delta \epsilon^\mu=-\frac{\theta}{\mpl}C^\mu\,.
\eeq
The explicit form of such $\epsilon$ is straightforward to find to the first nontrivial order in $\mpl^{-1}$, and is given by
\beq
&&\epsilon_0=-\frac{1}{4\mpl}\frac{1}{\laplacian}\hat{\pi}_{kk}+\mathcal{O}(\mpl^{-2})\,.\\
&&\epsilon_j=\frac{1}{\mpl}\frac{1}{\laplacian}\p_i\left( \hat{h}_{ij}-\frac{1}{2}\delta_{ij}\hat{h}_{kk} \right)+\mathcal{O}(\mpl^{-2})\,.~~
\eeq
The transformation properties can be readily verified using \eqref{hbrst} and \eqref{pibrst}. Therefore, a coherent state of the scalar field that satisfies the physicality condition of being annihilated by the BRST charge can be constructed as
\beq
\label{cohgravdress}
\ket{C}=e^{i \int d^3x \Pi_\phi^c(x) \hat{\phi}(x^\mu+\epsilon^\mu)}|\Omega\ra\,,
\eeq
with $\Pi_\phi^c(x)$ being an arbitrary function of the spacial coordinates and $x^0$ appearing in the argument of $\hat{\phi}$ sets the initial time (which we take to be at $x^0=0$). Notice that this state is not the most general one. In fact,  the initial expectation value of $\hat{\phi}$ vanishes. On the other hand, the configuration at hand does possess a nonzero kinetic energy. To be more specific, up to corrections of order $\mpl^{-2}$ the nontrivial initial conditions for one-point expectation values are as follows
\beq
&&\la C| \hat{\Pi}_\phi|C\ra (t=0)=\Pi_\phi^c+\mathcal{O}(\mpl^{-2})\,,\\
&&\laplacian \la C| \hat{h}_{ij}|C\ra (t=0)=\frac{\delta_{ij}}{8\mpl}{\Pi_\phi^c}^2+\mathcal{O}(\mpl^{-2})\,.~~~~~
\label{hfrompi}
\eeq
Notice that \eqref{hfrompi} is precisely the equation one would expect in Newtonian limit, for arbitrary $\Pi_\phi^c(x)$ while having $\phi_c(x)=0$. The generalization of this arguments for the BRST invariant dressing of $\hat{\Pi}_\phi$ is straightforward but the expression is cumbersome and will not be recited here.

\section{Non-linearities and Hubble scale} 

  So far we were discussing the de Sitter solution (\ref{hcl}) and its coherent state realization in theory of massless spin-$2$ with 
  a constant source. 
  This theory, both at classical and quantum levels, represents 
 a fully self-consistent limit of Einstein gravity,
\begin{eqnarray} \label{Limit} 
\mpl \, \rightarrow \, \infty\,,~~\, \lambda \, = \, {\rm finite}\,.
\end{eqnarray}
In this limit all non-linearities vanish.  
 Correspondingly, the solution (\ref{hcl}) as well as its 
 coherent state representation (\ref{CohH}) are exact. 
 
  This is fully consistent with the fact that the Hubble 
  scale  (\ref{Hubble}) is vanishing, despite that the cosmological constant source $\lambda$ is non-zero. Correspondingly, the effects of 
  de Sitter are experienced neither by the graviton nor by the external 
  particles coupled to it. For example, consider the  scalar with energy 
  momentum tensor $T^{\mu\nu}$ coupled to $\hat{h}{\mn}$   
  in (\ref{pertL}).  This scalar, while interacting with the coherent state
 of gravitons (\ref{CohH}),  effectively sees the following classical metric 
  \begin{eqnarray} \label{MetricEff} 
     g{\mn} \, = \,   \eta{\mn} \, + \, \frac{h{\mn}}{\mpl} \, + \,
  {\mathcal O}(\mpl^{-2}) \,,  
  \end{eqnarray}
where the classical field $h{\mn}$ is given by (\ref{hcl}). 
Obviously, because of the limit (\ref{Limit}), this is just a flat 
Minkowski metric. 

 Let us now assume that  $\mpl$ is taken finite.  This choice, of course,  makes $H$ non-zero.  In coordinates in question, significant departure from the Minkowski geometry is felt immediately near the Hubble radius and beyond. At much shorter radii, on the other hand, the probe scalar shall start feeling the non-trivial metric after a finite time. 
This will happen when $h{\mn}$ will become of order ${\mpl}$. 
As it is clear 
 from the solution (\ref{hcl}), the required time-scale for such a growth near the origin is given by the Hubble time, $t \sim H^{-1}$. 
    
    Obviously, the same applies to all non-linear self-interactions 
    of graviton. As it is well known, at tree-level 
  these interactions can be obtained by consistently coupling graviton to its own energy momentum tensor, order by order in $1/\mpl$ expansion.   
  The result fully coincides with the expansion of Einstein-Hilbert 
  action on Minkowski background.   
   In $n$-th order, the non-linear couplings exhibit the following 
    power-scaling with  $\mpl$,
    \begin{eqnarray} \label{NonlinearH}     
  \frac{h^n}{\mpl^n} \partial h \partial h\, .
    \end{eqnarray}
 The notation is of course highly schematic but suffices for making an important point. As we can see, just as in the case of 
 a probe scalar, the non-linearities become important 
 after the Hubble time.  Of course, neither the classical solution
 (\ref{hcl}) nor the corresponding quantum coherent  
 state (\ref{CohH}) are fit for adequately describing physics 
 beyond this time.  However, such long time-scales are beyond the concern of the present work. 
   
  Our main point is that BRST-invariant coherent state of gravitons 
 (\ref{CohH}) in linear theory of massless spin-$2$, 
 correctly captures the correlation between non-linearities 
 and the Hubble time.  This is a consistency check 
 of the presented description.  
 
Another consistency check is how the quantum depletion of the graviton coherent 
state captures the Gibbons-Hawking particle creation in de Sitter space.  This is what we shall consider next.    

\section{Gibbons-Hawking radiation in linear gravity} 

  The de Sitter spacetime exhibits the Gibbons-Hawking
 particle creation \cite{Gibbons:1977mu}. In a semiclassical treatment of de Sitter, this effect is described as a vacuum process.
  The coherent state picture offers a different way of looking 
  at the origin of this radiation.  In this description, de Sitter 
  is not a vacuum but rather an excited coherent state, constructed on top of the $S$-matrix vacuum of Minkowski. 
Therefore, the particle-creation represents a process of quantum decay of the graviton coherent state into different quanta. 
The rate of the decay \cite{Dvali:2013eja, Dvali:2017eba}, 
   \begin{eqnarray} \label{rate}     
  \Gamma \sim H \,, 
    \end{eqnarray}
and  the power of the emitted radiation,  
 \begin{eqnarray} \label{power}     
  P \,  \sim \, H^2 \,, 
    \end{eqnarray}
are in agreement with Gibbons-Hawking radiation of temperature 
$H$.  As argued in \cite{Dvali:2013eja, Dvali:2017eba}, due to backreaction, this decay inevitably leads to a loss of coherence and generation of entanglement. This results in a complete departure from the classical picture after a half-decay. This time scale is called quantum break-time.

Can the glimpses of the above process be 
captured by the coherent state (\ref{CohH}) of  linearized theory?  
  The effect can be read out in two different ways. 
 The first way is by analysing the equation for the probe scalar in the background metric (\ref{MetricEff}), with the solution 
 (\ref{hcl}).  Of course, this must be understood as the 
 leading order result in the expansion in $1/\mpl$. 
 During the computation the scale $\mpl$ must be kept finite. 
 The limit (\ref{Limit}) can be taken afterwords.     

  This approximation gives the  particle creation rate
 (\ref{rate}) and the emission power (\ref{power}).  This indicates that the state 
 produces on average one particle of energy $\sim H$ 
per time $t\sim H^{-1}$. Since we are not going 
beyond this time, the computation is informative, at least,  order of 
magnitude wise.   
 
 The above also indicates that within the exact validity of linear 
 theory, the particle production vanishes, since 
 $H$ vanishes in the limit (\ref{Limit}). 
Remarkably, there is a way of  capturing a  non-zero particle-creation rate even in this limit.
  
 It has been shown in \cite{Dvali:2021bsy} that by taking a 
 so-called ``species limit'', one can ensure that the linear gravity represents an exact description of full quantum theory, while simultaneously maintaining the collective phenomena, such as particle creation in de Sitter.  This setup is ready-made for our purposes.  Therefore, below, we shall follow 
the construction of \cite{Dvali:2021bsy} and shall implement our de Sitter coherent state 
(\ref{CohH}) within this framework. This shall allow us to extract
the Gibbons-Hawking particle creation process in a  
BRST-invariant coherent state description of de Sitter.

 The idea of the species limit is the following. 
 Let us, instead of a single (massless) scalar, introduce 
 $N$ copies of them $\phi_j$ where 
 $j = 1,2, ..., N$ is the species index.  
 For definiteness, we assume no self-interactions 
 for scalars. 
 
 The introduction of 
 species has the effect of lowering the gravitational cutoff 
 $\Lambda_{\rm gr}$  \cite{Dvali:2007hz,Dvali:2007wp}, to
   \begin{eqnarray} \label{cutoff}     
   \Lambda_{\rm gr} \, = \, \frac{\mpl}{\sqrt{N}} \,. 
    \end{eqnarray}
Now, the species limit is defined as 
\begin{eqnarray} \label{SpLimit}     
   \mpl \, \rightarrow \, \infty \,,~ \, N \, \rightarrow \, \infty\,, ~
 \, \Lambda_{\rm gr} \, = \, {\rm finite}\, . 
    \end{eqnarray}
    This limit is somewhat analogous to the 't Hooft's planar limit 
    in $SU(N)$ QCD \cite{tHooft:1973alw}, in the following sense. 
 The quantum gravitational coupling, which at any finite energy scale $q$ is defined as   
    \begin{eqnarray} \label{alphaGR}     
  \alpha_{\rm gr} \, \equiv  \, \frac{q^2}{\mpl^2} \,,  
    \end{eqnarray}
vanishes in the limit (\ref{SpLimit})  as 
$1/N$. Correspondingly, all quantum gravitational 
processes in which $\alpha_{\rm gr}$ is not accompanied by $N$, vanish.  In certain sense, this gives 
a greater simplification than the 't Hooft limit in QCD. This is  due to the fact that graviton, unlike gluons in QCD,  carries no species index.  

 At the loop level, this implies 
vanishing of all loop contributions except the renormalization of the graviton kinetic term, which is
resummable.   Of course, all non-linear self-interactions of 
the graviton vanish. The resulting theory is the 
linearized gravity coupled to $N$ scalar species, 
\begin{equation}
\label{pertLSP}
	\mathcal{L}=\mathcal{L}_\phi+\mathcal{L}_h-\frac{1}{2\mpl}\hat{h}_{\mu\nu}\sum_{j=1}^N\hat{T}_j^{\mu\nu} \,.
\end{equation}
Notice, the limit (\ref{SpLimit}) ensures that 
the above form provides an exact description on any
state in which the occupation number of quanta 
increases slower than $N$.  
Of course,  the coupling of graviton to each particular species is zero but 
the collective effect is non-vanishing.  Due to this, the coupling 
between graviton and species must be kept in the
Lagrangian, even in the species limit. 

  Next, we shall add a constant source $\lambda$ and keep it 
  non-zero and finite while simultaneously taking the species limit 
  (\ref{SpLimit}). The resulting 
 state represents a linearized de Sitter coherent state 
  (\ref{CohH}) interacting with $N$ particle species. 
 The difference with the case of a single scalar is that 
 the rate of the particle-creation is  enhanced by the factor 
 $N$,  
   \begin{eqnarray} \label{rateN}     
  \Gamma \sim H \, N \,. 
    \end{eqnarray}
Of course, in the present limit $\Gamma$ diverges due to the infinite Hubble  volume. 
Despite this,  
 the particle production rate per unit volume, which is given by 
  \begin{eqnarray} \label{HN}     
\frac{\Gamma}{V}\sim H^4N=\left( \frac{\lambda}{\Lambda_{\rm gr}} \right)^2\,, 
    \end{eqnarray}
is finite.

 From the quantum corpuscular point of view, 
 the rate (\ref{rateN}) has a clear physical meaning. 
 The particle creation comes from the re-scattering of the
 constituent gravitons into the scalar species. 
 Consider for example, a scattering of two gravitons into 
 a pair of scalars.  The rate of the process is given by 
   \begin{eqnarray} \label{rateNC}     
  \Gamma \sim H \alpha_{\rm gr}^2 N_{\rm gr}^2 \, N \,,  
    \end{eqnarray}
where $\alpha_{\rm gr} = H^2/\mpl^2$ is the coupling 
between the coherent state gravitons and scalars and 
$N_{\rm gr} = \mpl^2/H^2$ is the occupation number of 
gravitons in the coherent state per Hubble volume. The two exactly compensate 
each other,  $\alpha_{\rm gr} N_{\rm gr} = 1$. What remains is the enhancement by an infinite factor $N$. 
The resulting particle-creation rate per unit volume (\ref{HN})
is finite.  
 
 We see that the species limit (\ref{SpLimit}) allows for  
the computation of the  
Gibbons-Hawking radiation from the graviton coherent state 
(\ref{CohH}).  
In this limit, the effect is finite,  despite the fact that the Gibbons-Hawking temperature is zero and the Hubble time is infinite.  
This is a particular manifestation of a general phenomenon 
that species magnify the effects of quantum gravity.  

  Notice, as another check of consistency of our coherent state description, the  back-reaction on the coherent state  from the particle-creation vanishes despite the finite production rate per unit volume. This is due to the fact that  in the limit 
 (\ref{SpLimit}) the mean number density of the constituents is infinite and their frequencies ($\sim H$) are zero. This results in infinite quantum break-time, since in species regime the latter scales as \cite{Dvali:2020etd, Dvali:2021bsy}, 
 \beq \label{TQ}     
  t_Q \, \sim \, \frac{\mpl^2}{H^3 N} \, \sim \,
  \frac{\Lambda_{\rm gr}^2}{H^3} \,, 
    \eeq
 which in the present case is infinite. 
 The time-scale for 
  the half-decay of the coherent state is of the same order.  This aspect is 
  also correctly captured by the linear theory.

\section{Summary} 

 The $S$-matrix formulation of quantum gravity excludes 
 de Sitter vacua \cite{Dvali:2020etd}.  The only remaining option for realzing de Sitter in quantum gravity is by representing it as an excited  
 state on top of a valid $S$-matrix vacuum, such as Minkowski. 
 Since the state must be close to classical, the natural candidate
 for it is a coherent state.   
 
  This approach, originally adopted in \cite{Dvali:2013eja,
Dvali:2014gua, Dvali:2017eba},  offers a microscopic understanding of known de Sitter phenomena.  At the same time, it reveals the new effects that 
 in ordinary semiclassical treatment are not visible. 
 The processes such as Gibbons-Hawking radiation, which 
 in ordinary picture are viewed as vacuum processes, 
 in the $S$-matrix picture are described as actual decay of the 
 graviton coherent state. This makes it clear that 
 at finite $\mpl$, the de Sitter is subjected to backreaction. 
 This backreaction, limits the duration 
 of the classical de Sitter phase by its quantum break-time, $t_Q$. 
 For finite values of  $\mpl$ and $H$, this time is finite, thereby 
 eliminating possibility of the eternal de Sitter cosmology.
 This has important consequences both for cosmology 
 and for particle physics. 

 The above gives a fundamental importance to understanding the viability of consistent formulation of de Sitter coherent state.  In the present paper we have offered a BRST invariant construction of such a state. 

  However, we  approached the issue through  a prism of a broader question. This is a BRST invariant formulation of states produced by classical sources in quantum gauge theories.  
 In this work, we focused on Abelian gauge theories such as QED and  linear Einstein gravity.
   
   First, starting with quantum electrodynamics, we have discussed the coherent state description of the charged scalar field configuration. 
We have shown that, in order to comply with  physicality constraints,  charges must be dressed in a 
special way. Namely, the non-asymptotic coherent state of the charged scalar field cannot be made BRST invariant by dressing it with the coherent state of the electromagnetic field, even though coherent states of photons on their own are fully consistent with the BRST condition. Instead, we build matter coherent states out of gauge invariant operators \`a la Dirac \cite{Dirac:1955uv}. 
 Next we have studied the coherent state of photons produced 
 in the presence of external classical charges.  
 
 An interesting toy example that shares certain qualitative properties with de Sitter in gravity can be set up by introducing classical and quantum charges simultaneously. The role of the cosmological constant is played by the uniform source of constant charge density. In the absence of dynamical quantum charges, such 
 a source of infinite extent produces a linearly growing 
 (in space-time coordinates) electric field, or  
 equivalently, a  quadratically growing vector potential $A_{\mu}$. 
 This is similar to the graviton field produced by the cosmological constant in linearized Einstein gravity.
 
 Introduction of quantum charges with finite couplings to
 corresponding gauge fields (photon in QED, graviton in Einstein)    
 tames the unbounded growth in both cases. 
 In QED this happens through the Schwinger 
 pair-creation which reduces the electric field below the threshold.  
   In gravity, the growth is tamed by couplings of graviton
  to itself and to other species. These become important on the Hubble scale.  In addition, the Gibbons-Hawking particle creation 
  shares some similarity with Schwinger discharge.   
    
   In gravity, we have found that the de Sitter space can be constructed as a BRST invariant coherent state of gravitons on top of the Minkowski space within the Gaussian theory of massless spin-2 field. We have also introduced the gravitating scalar matter and have shown how to dress its coherent state perturbatively in $\mpl^{-1}$.
   
   The introduction of $N$ species of scalars
  allows to implement our coherent state construction 
 in the species limit (\ref{SpLimit}), which promotes the linear gravity coupled to $N$ species into an exact description.
  The only quantum gravitational effects surviving in this 
  limited are the collective phenomena in which the 
powers of $1/\mpl^2$  are  compensated by 
  corresponding  powers of $N$. 
   An important example of a non-zero collective effect
  surviving in the species limit  is the Gibbons-Hawking radiation.
    Correspondingly,  the species limit allows us to observe that     
 BRST invariant coherent state, 
  describing the de Sitter in linearized gravity, exhibits 
 features of full non-linear de Sitter.  
 
\section*{\textit{Note added} }
After submission of this work, it was brought to our attention that the gauge invariant observable $\phi(x+\epsilon)$, discussed in section \ref{scalarsection}, has been previously studied in a different context \cite{Donnelly:2015hta,Frob:2017gyj}.
  
  \section*{ACKNOWLEDGEMENTS}
  This work was supported in part by the Humboldt Foundation under Humboldt Professorship Award, by the Deutsche Forschungsgemeinschaft (DFG, German Research Foundation) under Germany's Excellence Strategy - EXC- 2111 - 390814868, and Germany's Excellence Strategy under Excellence Cluster Origins.

\end{document}